\documentclass[aps,prl,twoside,twocolumn,superscriptaddress,floatfix,nofootinbib,showpacs]{revtex4}
\usepackage{amsmath}
\usepackage{bm}
\usepackage{xcolor}

\newcommand\beq{\begin{equation}}
\newcommand\eeq{\end{equation}}
\newcommand\beqn{\begin{eqnarray}}
\newcommand\eeqn{\end{eqnarray}}

\newcommand{\ba}{\begin{eqnarray}}
\newcommand{\ea}{\end{eqnarray}}
\newcommand{\be}{\begin{equation}}
\newcommand{\ee}{\end{equation}}

\newcommand\lsim{\mathrel{\rlap{\lower4pt\hbox{\hskip1pt$\sim$}}
        \raise1pt\hbox{$<$}}}
\newcommand\gsim{\mathrel{\rlap{\lower4pt\hbox{\hskip1pt$\sim$}}
        \raise1pt\hbox{$>$}}}

\newcommand{\jcap}{{J.~Cosm.~Astrop.~Phys.}}
\newcommand{\araa}{{Annu.~Rev.~Astron.~Astrophys.}}
\newcommand{\aap}{{Astron.~Astrophys.}}
\newcommand{\apjl}{{Astrophys.~J.~Lett.}}
\newcommand{\apjs}{{Astrophys.~J.~Supp.}}
\newcommand{\aj}{{Astron.~J.}}
\newcommand{\mnras}{{Mon.~Not.~R.~Astron.~Soc.}}
\usepackage{graphicx}
\usepackage[large]{subfigure}
\usepackage{amssymb, amsmath}
\usepackage[amssymb]{SIunits}
\usepackage{epstopdf}
\usepackage{aas_macros}
\usepackage{natbib}

\begin{document}

\title{The Kinematic Sunyaev-Zel'dovich Effect with Projected Fields: A Novel Probe of the Baryon Distribution with \emph{Planck}, \emph{WMAP}, and \emph{WISE} Data}
%%%%%%%%%%%%%%%%%%%%%%%%%%%%%%%%%%%%%%%%%%%%%%%%%%%%%%%%%%%%%%%%%%%%%%%%%%%
 \author{J.~Colin~Hill}\email{jch@astro.columbia.edu}
 \affiliation{Dept.~of Astronomy, Pupin Hall, Columbia University, New York, NY USA 10027}
 \author{Simone~Ferraro}
 \affiliation{Miller Institute for Basic Research in Science, University of California, Berkeley, CA, 94720, USA}
  \author{Nick~Battaglia}
 \affiliation{Dept.~of Astrophysical Sciences, Peyton Hall, Princeton University, Princeton, NJ USA 08544}
 \author{Jia~Liu}
 \affiliation{Dept.~of Astronomy, Pupin Hall, Columbia University, New York, NY USA 10027} 
 \author{David~N.~Spergel}
 \affiliation{Dept.~of Astrophysical Sciences, Peyton Hall, Princeton University, Princeton, NJ USA 08544}
%%%%%%%%%%%%%%%%%%%%%%%%%%%%%%%%%%%%%%%%%%%%%%%%%%%%%%%%%%%%%%%%%%%%%%%%%%%
\begin{abstract}
The kinematic Sunyaev-Zel'dovich (kSZ) effect --- the Doppler boosting of cosmic microwave background (CMB) photons due to Compton-scattering off free electrons with non-zero bulk velocity --- probes the abundance and distribution of baryons in the Universe.  All kSZ measurements to date have explicitly required spectroscopic redshifts.  Here, we implement a novel estimator for the kSZ -- large-scale structure cross-correlation based on projected fields: it does not require redshift estimates for individual objects, allowing kSZ measurements from large-scale imaging surveys.  We apply this estimator to cleaned CMB temperature maps constructed from \emph{Planck} and \emph{Wilkinson Microwave Anisotropy Probe} (\emph{WMAP}) data and a galaxy sample from the \emph{Wide-field Infrared Survey Explorer} (\emph{WISE}).  We measure the kSZ effect at $3.8$--$4.5\sigma$ significance, depending on the use of additional \emph{WISE} galaxy bias constraints.  We verify that our measurements are robust to possible dust emission from the \emph{WISE} galaxies.  Assuming the standard $\Lambda$CDM cosmology, we directly constrain $\left( {f_{b}}/{0.158} \right) \left( {f_{\rm free}}/{1.0} \right) = 1.48 \pm 0.19$ (statistical error only) at redshift $z \approx 0.4$, where $f_{b}$ is the fraction of matter in baryonic form and $f_{\rm free}$ is the free electron fraction.  This is the tightest kSZ-derived constraint reported to date on these parameters.  Astronomers have long known that baryons do not trace dark matter on $\sim$kpc scales and there has been strong evidence that galaxies are baryon-poor.  The consistency between the $f_{b}$ value found here and the values inferred from analyses of the primordial CMB and Big Bang nucleosynthesis (BBN) verifies that baryons approximately trace the dark matter distribution down to $\sim$Mpc scales.  While our projected-field estimator is already competitive with other kSZ approaches when applied to current datasets (because we are able to use the full-sky \emph{WISE} photometric survey), it will yield enormous signal-to-noise when applied to upcoming high-resolution, multi-frequency CMB surveys.
\end{abstract}
\maketitle

%%%%%%%%%%%%%%%%%%%%%%%%%%%%%%%%%%%%%%%%%%%%%%
\emph{Introduction---}
In the standard cosmological paradigm, the primordial CMB anisotropies~\cite{Spergeletal2003,Hinshawetal2013,Planck2015params} and the abundance of light elements formed in the primordial plasma (BBN)~\cite{Steigman2007} imply that the baryon density ($\rho_b$) is approximately one-sixth of the total matter density ($\rho_m$): $f_{b} \equiv \rho_{b}/\rho_{m} \approx 0.158$~\cite{Planck2015params}.  Yet, in galaxies, $f_b$ is a factor $\approx 2$--$3$ lower~\cite{Fukugitaetal1998,Bregman2007}.  The ``missing baryons'' are thought to reside in an ionized, diffuse, warm-hot plasma.  They have been difficult to detect in X-ray emission or absorption line studies, although recent \emph{Hubble Space Telescope} observations indicate many such baryons are indeed located in the circumgalactic medium of typical low-redshift galaxies ($z \approx 0.2$)~\cite{Werketal2014}.

In this \emph{Letter}, we measure the cross-correlation of the kinematic Sunyaev-Zel'dovich (kSZ) effect and the distribution of infrared-selected galaxies using a novel estimator.  The inferred $f_b$ is consistent with the cosmological value, implying that most of the baryons are in ionized gas tracing the dark matter on $\gtrsim$Mpc scales.

The kSZ effect is the Doppler boosting of CMB photons as they Compton-scatter off free electrons moving with a non-zero line-of-sight (LOS) velocity in the CMB rest frame~\cite{Sunyaev-Zeldovich1972,Sunyaev-Zeldovich1980,Ostriker-Vishniac1986}.  This effect leads to a shift in the observed CMB temperature, with an amplitude proportional to the mass in electrons and their LOS velocity, which is equally likely to be positive or negative.  Since the electrons and dark matter are expected to follow the same large-scale velocity field, the kSZ signal traces the overall mass distribution, unlike the thermal Sunyaev-Zel'dovich (tSZ) effect, whose amplitude is proportional to the electron pressure and is thus primarily sourced by galaxy groups and clusters~\cite{Komatsu-Seljak2002,Tracetal2011,BBPS2012b,BHM2015}.

The kSZ effect was first detected using data from the \emph{Atacama Cosmology Telescope} (\emph{ACT})~\cite{Handetal2012} by studying the pairwise momenta of galaxy groups and clusters~(e.g.,~\cite{Ferreiraetal1999}).  Subsequent detections using this approach were reported by \emph{Planck}~\cite{Planck2016kSZ,Planck2016kSZinterp} and the \emph{South Pole Telescope}~\cite{Soergeletal2016}.  Measurements using a velocity-field reconstruction estimator~\cite{Hoetal2009,Lietal2014} were reported by \emph{Planck}~\cite{Planck2016kSZ,Planck2016kSZinterp} and \emph{ACTPol}~\cite{Schaanetal2016}.\footnote{A high-resolution analysis of an individual galaxy cluster also found evidence for the kSZ effect~\cite{Mroczkowskietal2012,Sayersetal2013}.}  Crucially, these estimators both rely on spectroscopic data, which are more expensive and time-consuming to acquire than photometric imaging data.  Even excellent photometric redshifts yield significantly decreased signal-to-noise ($S/N$) for these estimators, compared to spectroscopic data~\cite{Keisler-Schmidt2013,Flenderetal2015,Soergeletal2016}.

We implement a kSZ estimator based on projected fields: it can be applied to any large-scale structure sample, including galaxies, quasars, or gravitational lensing maps, allowing analyses of densely-sampled, full-sky surveys.  First suggested in Refs.~\cite{Doreetal2004,DeDeoetal2005}, the estimator relies on the fact that a frequency-cleaned CMB temperature map contains kSZ information on small scales, regardless of whether external velocity information is available.  The kSZ information can be accessed in cross-correlation with tracers of the large-scale density field.  To avoid the cancellation of equally-likely positive and negative kSZ signals, the temperature map is squared in real space before cross-correlating.

We measure the kSZ${}^2$--tracer cross-correlation in this \emph{Letter} using data from \emph{Planck}~\cite{Planck2015overview}, \emph{WMAP}~\cite{Bennettetal2013}, and \emph{WISE}~\cite{Wrightetal2010}.  On large scales, this method constrains $f_{b}$ or $f_{\rm free}$ without the need for individual halo mass estimates.  Here, $f_{\rm free}$ denotes the fraction of electrons that are not bound in neutral media, and thus take part in Compton scattering.  In a companion paper~\cite{Ferraroetal2016} (hereafter F16), we provide theoretical details, compare to numerical simulations, and investigate the reach of this method for upcoming surveys.  We assume the best-fit \emph{Planck} $\Lambda$CDM cosmological parameters~\cite{Planck2015params}.
%%%%%%%%%%%%%%%%%%%%%%%%%%%%%%%%%%%%%%%%%%%%%%

%%%%%%%%%%%%%%%%%%%%%%%%%%%%%%%%%%%%%%%%%%%%%%
\emph{Theory---}
The kSZ-induced fractional CMB temperature shift, $\Theta^{\rm kSZ}(\hat{\mathbf{n}}) \equiv \Delta T^{\rm kSZ}(\hat{\mathbf{n}})/T_{\rm CMB}$, in a direction $\hat{\mathbf{n}}$ on the sky is
\be
\Theta^{\rm kSZ}(\hat{\mathbf{n}}) = - \frac{1}{c} \int_0^{\eta_{\rm re}} d\eta \, g(\eta) \, \mathbf{p}_e \cdot \mathbf{\hat{n}} \,,
\label{eq.kSZdef}
\ee
where $T_{\rm CMB}$ is the mean CMB temperature, $\eta(z)$ is the comoving distance to redshift $z$, $\eta_{\rm re}$ is the comoving distance to the end of hydrogen reionization, $g(\eta) = e^{-\tau} d\tau/d\eta$ is the visibility function, $\tau$ is the optical depth to Thomson scattering, and $\mathbf{p}_e = (1+\delta_e) \mathbf{v}_e$ is the electron momentum, with $\delta_e \equiv (n_e - \bar{n}_e)/\bar{n}_e$ the electron overdensity, $n_e$ the free electron number density, and $\mathbf{v}_e$ the electron peculiar velocity.

The projected galaxy overdensity is
\be
\delta_g(\hat{\mathbf{n}}) = \int_0^{\eta_{\rm max}} d\eta \, W_g(\eta) \, \delta_m(\eta \hat{\mathbf{n}}, \eta) \,,
\label{eq.deltagdef}
\ee
where $\eta_{\rm max}$ is the maximum comoving distance of the galaxy sample, $\delta_m \equiv (\rho_m - \bar{\rho}_m)/\bar{\rho}_m$ is the matter overdensity, $\rho_m$ is the matter density, and $W_g(\eta) = b_g p_s(\eta)$ is the projection kernel.  Here, $b_g$ is the linear galaxy bias and $p_s(\eta)$ is the distribution of source galaxies, normalized to have unit integral.

To downweight angular scales dominated by primary CMB fluctuations and detector noise, we apply a filter $F_{\ell}$ in harmonic space~\cite{Doreetal2004}, $F_{\ell} = C_{\ell}^{\rm kSZ} / C_{\ell}^{\rm tot}$, where $C_{\ell}^{\rm kSZ}$ is the (theoretical) kSZ power spectrum and $C_{\ell}^{\rm tot}$ is the total fluctuation power, which includes the primary CMB, kSZ effect, integrated Sachs-Wolfe (ISW) effect, noise, and residual foregrounds.  Our theoretical kSZ power spectrum template is derived from cosmological hydrodynamics simulations~\cite{BBPSS2010} and semi-analytic models~\cite{Battagliaetal2013reionC}.  F16 shows $F_{\ell}$.  The telescope beam window function is modeled as an additional filter (here, a Gaussian with FWHM $= 5$ arcmin).

Direct cross-correlation of $\Theta^{\rm kSZ}$ and $\delta_g$ is expected to vanish due to the $\mathbf{v}_e \rightarrow -\mathbf{v}_e$ symmetry of the kSZ signal.  We thus square the filtered $\Theta^{\rm kSZ}$ in real space before cross-correlating with $\delta_g$~\cite{Doreetal2004,DeDeoetal2005}.  In the Limber approximation~\cite{Limber1953}, this angular cross-power spectrum is
\be
C_\ell^{\rm{kSZ}^2 \times \delta_g} = \frac{1}{c^2} \int_0^{\eta_{\rm max}} \frac{d \eta} {\eta^2} W_{g}(\eta) g^2(\eta) \mathcal{T}\left(k = \frac{\ell}{\eta}, \eta\right) \,,
\label{eq.kSZ2deltagPS}
\ee
where $\mathcal{T}(k, \eta)$ is an integral over the hybrid bispectrum $B_{\delta p_{\hat{\mathbf{n}}} p_{\hat{\mathbf{n}}} }$ of one density contrast $\delta$ and two LOS electron momenta $p^e_{\hat{\mathbf{n}}}$.  We approximate $B_{\delta p_{\hat{\mathbf{n}}} p_{\hat{\mathbf{n}}} } \approx v_{\rm rms}^2 B_m^{\rm NL}/3$~\cite{Doreetal2004,DeDeoetal2005}, where $v_{\rm rms}^2$ is the linear-theory velocity dispersion and $B_m^{\rm NL}$ is the non-linear density bispectrum, for which we use a fitting function from numerical simulations~\cite{Gil-Marinetal2012}.  We cross-check this approach with hydrodynamical simulations in F16.

The visibility function $g(\eta) \propto f_{b} f_{\rm free}$, and thus $C_\ell^{\rm{kSZ}^2 \times \delta_g} \propto f_{b}^2 f_{\rm free}^2$, weighted by the kernels in Eq.~(\ref{eq.kSZ2deltagPS}).  At the current $S/N$ level, we cannot constrain the redshift dependence of these quantities, and thus simply fit an overall amplitude.  We also assume that the free electrons trace the overall density field on the scales accessible to \emph{Planck}; higher-resolution experiments can directly measure the free electron profiles around galaxies and clusters, which influence the small-scale shape of $C_\ell^{\rm{kSZ}^2 \times \delta_g}$.

Since $C_\ell^{\rm{kSZ}^2 \times \delta_g}$ is quadratic in the CMB temperature, this estimator receives a contribution from gravitational lensing of the CMB~(e.g.,~\cite{Lewis-Challinor2006}).  We compute this term at first order in the lensing potential and cross-validate with simulations~\cite{Sehgaletal2010} in F16.  The lensing contribution is proportional to $b_g$.  We can thus improve the $C_\ell^{\rm{kSZ}^2 \times \delta_g}$ measurement by externally constraining $b_g$ via cross-correlation of the \emph{WISE} galaxies with \emph{Planck} CMB lensing maps ($C_\ell^{\kappa_{\rm CMB} \delta_g}$, where $\kappa_{\rm CMB}$ is the lensing convergence).  Alternatively, the lensing contamination can be fit simultaneously with the kSZ${}^2$ amplitude and marginalized over.
%%%%%%%%%%%%%%%%%%%%%%%%%%%%%%%%%%%%%%%%%%%%%%

%%%%%%%%%%%%%%%%%%%%%%%%%%%%%%%%%%%%%%%%%%%%%%
\emph{Data---}
The kSZ signal is extracted from a frequency-cleaned CMB temperature map~\cite{Bobinetal2015},\footnote{{\tt http://www.cosmostat.org/research/cmb/planck\char`_wpr2}} which we further clean as described below.   This map is constructed from a joint analysis of the nine-year \emph{WMAP}~\cite{Bennettetal2013} and \emph{Planck} full mission~\cite{Planck2015overview} full-sky temperature maps.  The CMB is separated from other components in the microwave sky using ``local-generalized morphological component analysis'' (LGMCA), a technique relying on the sparse distribution of non-CMB foregrounds in the wavelet domain~\cite{Bobinetal2013,Bobinetal2014}.  The method reconstructs a full-sky CMB map with minimal dust contamination.  Importantly for our purposes, the tSZ signal is explicitly removed in the map construction (unlike in, e.g., the \emph{Planck} SEVEM, NILC, or SMICA component-separated CMB maps~\cite{Planck2015component}).  Components that preserve the CMB blackbody spectrum are not removed, including the kSZ and ISW signals.

We isolate the kSZ signal in the LGMCA temperature map using a filter, as described above.  We set $C_{\ell}^{\rm tot} = \hat{C}_{\ell}^{\rm LGMCA}$, where $\hat{C}_{\ell}^{\rm LGMCA}$ is the measured power spectrum of the LGMCA map.  We further set the filter to zero at $\ell \lsim 100$ and $\ell \gsim 3000$ in order to remove ISW contamination and noise-dominated modes, respectively.  We multiply by appropriate hyperbolic tangent functions at these boundaries to allow the filter to smoothly interpolate to zero, and normalize the filter such that its maximum value is unity, which occurs at $\ell \approx 2200$.

We construct a galaxy sample from \emph{WISE}, which imaged the sky in four photometric bands between 3.4 and 22 $\mu {\rm m}$.  Our color-based selection criteria match Ref.~\cite{Ferraroetal2014}, originally based on Ref.~\cite{2011ApJ...735..112J}.  The redshift distribution of these galaxies peaks at $z \approx 0.3$ and extends to $z \approx 1$ ($\langle z \rangle \approx 0.4$)~\cite{2013AJ....145...55Y}.  Their luminosities are similar to that of the Milky Way ($L \sim L^*$).  We apply a Galactic mask to remove stellar contamination (which would not bias our results, but only add to the noise).  We combine this mask with a point-source mask removing all sources detected at $>5 \sigma$ in the \emph{Planck} data~\cite{Hill-Spergel2014}, leaving a sky fraction $f_{\rm sky} = 0.447$ and 46.2 million \emph{WISE} galaxies.

We use the 2013 and 2015 \emph{Planck} CMB lensing maps~\cite{Planck2013lensing,Planck2015lensing} to place external constraints on the \emph{WISE} galaxy bias $b_g$.
%%%%%%%%%%%%%%%%%%%%%%%%%%%%%%%%%%%%%%%%%%%%%%

%%%%%%%%%%%%%%%%%%%%%%%%%%%%%%%%%%%%%%%%%%%%%%
\emph{Analysis---}
We apply the $C_\ell^{\rm{kSZ}^2 \times \delta_g}$ estimator described above to the filtered LGMCA temperature map and the \emph{WISE} galaxy density map.  Although the LGMCA map already shows very little dust contamination, we explicitly remove any dust associated with the \emph{WISE} galaxies by determining $\alpha$ that nulls the cross-correlation of $\delta_g$ and $((1+\alpha)T_{\rm LGMCA} - \alpha T_{\rm dust})$, where $T_{\rm dust}$ is a dust template constructed from a CMB-free combination of the filtered \emph{Planck} 217 and 545 GHz maps~\cite{SFH2015}: $T_{\rm dust} = 1.0085 (T_{545} - T_{217})$.  We find $\alpha_{\rm min} = -0.0002 \pm 0.0001$ and subsequently construct $T_{\rm clean} = (1+\alpha_{\rm min})T_{\rm LGMCA} - \alpha_{\rm min}T_{\rm dust}$.  All results are nearly identical whether we use $T_{\rm clean}$, the original $T_{\rm LGMCA}$, or a version of $T_{\rm clean}$ constructed using the \emph{Planck} 857 GHz map as a dust template.

\begin{figure}
\centering
\includegraphics[width=0.5\textwidth]{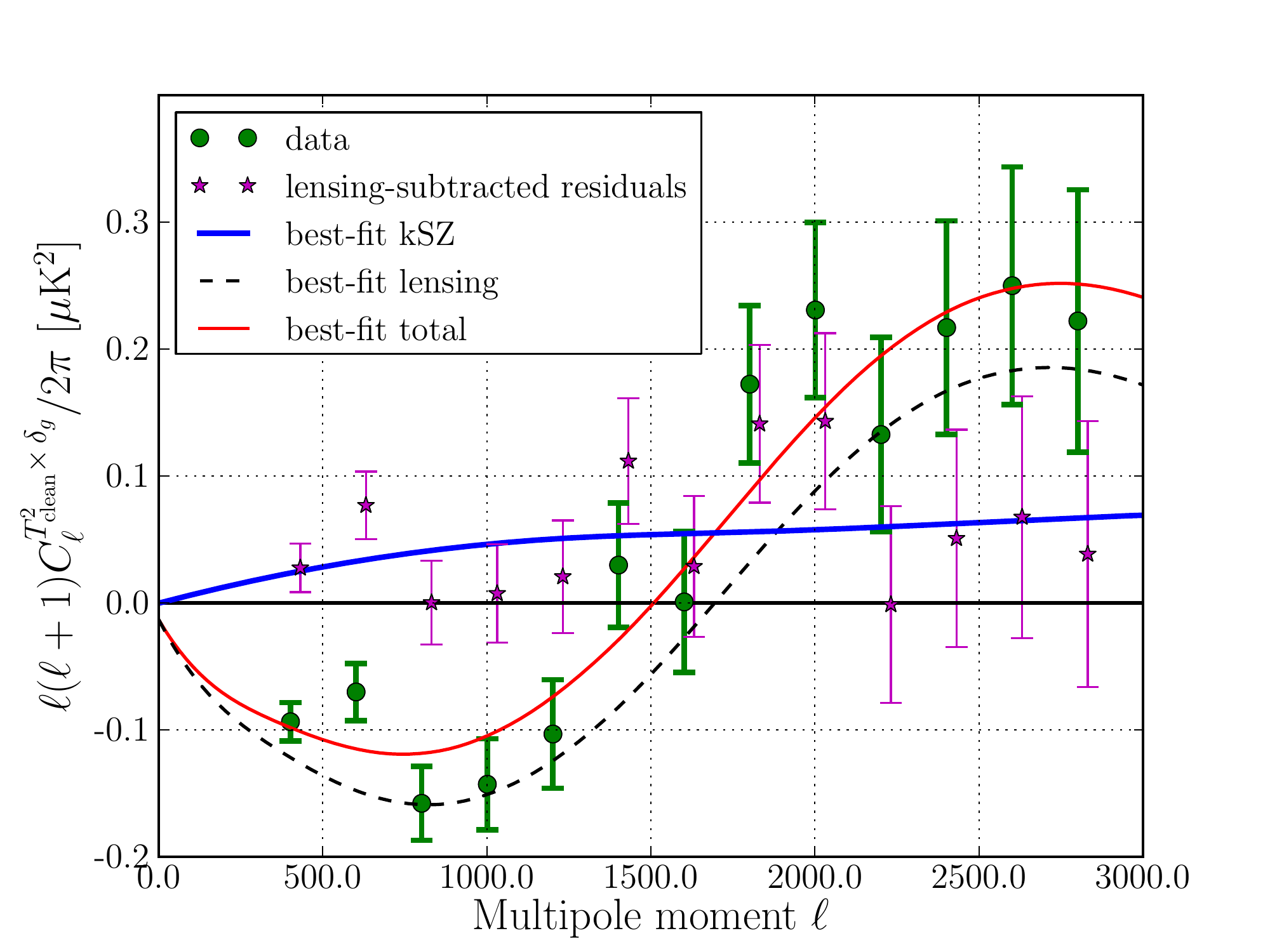}
\caption{Cross-power spectrum of filtered, squared $T_{\rm clean}$ map with \emph{WISE} galaxies (green circles).  The thick blue, dashed black, and thin red curves show the best-fit kSZ, CMB lensing, and kSZ $+$ lensing power spectra, respectively.  The kSZ signal is detected at $3.8\sigma$ significance.  No external galaxy bias constraint is used in these fits; including a prior from cross-correlating the \emph{WISE} galaxies with \emph{Planck} CMB lensing maps produces nearly identical best-fit results (see Table~\ref{tab:fits}), with a kSZ significance of $4$--$4.5\sigma$.  For visual purposes only, the magenta squares show the residual excess left after subtracting the best-fit lensing template from the data points.
\label{fig.kSZ2xWISE}}
\end{figure}

Fig.~\ref{fig.kSZ2xWISE} shows our measurement of $C_{\ell}^{T_{\rm clean}^2 \times \delta_g}$.  The dominant oscillatory shape is due to the CMB lensing contribution.  We measure the signal in thirteen linearly-spaced multipole bins between $\ell = 300$ and $\ell = 2900$ (bin width $\Delta \ell = 200$).  The lower multipole limit avoids the ISW signal.  We correct for the effects of the mask (apodized with a Gaussian taper of FWHM $= 10$ arcmin) using standard methods~\cite{MASTER} (the beam window function is forward-modeled in the theory calculations as described above).  Error bars on the cross-power spectrum are estimated in the Gaussian approximation using the measured auto-spectra of the $T_{\rm clean}^2$ and $\delta_g$ maps.

\begin{figure}
\centering
\includegraphics[width=0.5\textwidth]{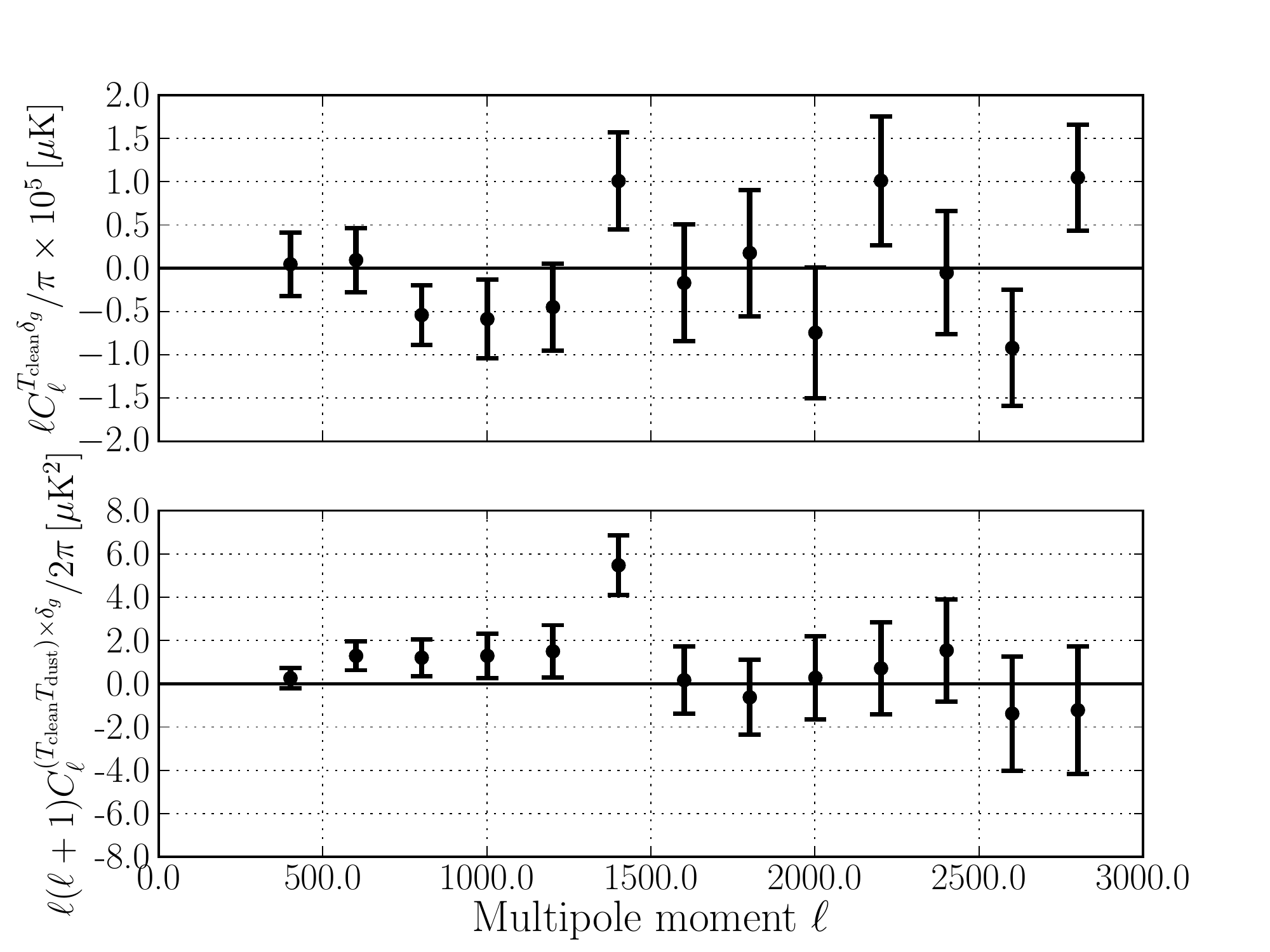}
\caption{Dust null tests.  \emph{Top}: Cross-correlation of $T_{\rm clean}$ with \emph{WISE} galaxies.  This verifies that any mean emission (e.g., dust, radio, or tSZ) of the galaxies is removed in $T_{\rm clean}$.  \emph{Bottom}: Cross-correlation of $(T_{\rm clean} T_{\rm dust})$ with \emph{WISE} galaxies.  This further verifies that any \emph{WISE}-galaxy-correlated dust emission in $T_{\rm clean}$ is sufficiently removed.  Re-scaling $T_{\rm dust}$ from $545$ GHz to $100$--$217$ GHz (a factor of $\approx 400$--$500$) yields a dust contribution to the data points in Fig.~\ref{fig.kSZ2xWISE} of $\lesssim 0.003 \, \mu{\rm K}^2$, well below the statistical errors.
\label{fig.nulltests}}
\end{figure}

We consider several tests to ensure that our measurement is not an artifact or due to contamination.  We process an LGMCA noise map ($T_{\rm noise}$) constructed from the half-difference of splits of the \emph{Planck} and \emph{WMAP} data.  The resulting cross-correlation of $T_{\rm noise}$ with $\delta_g$ is consistent with null (probability-to-exceed $p = 0.63$), as is $T_{\rm noise}^2$ with $\delta_g$ ($p = 0.27$).  We can consider the cross-correlation of $T_{\rm clean}$ (not squared) with $\delta_g$ as a null test for the mean dust contamination (Fig.~\ref{fig.nulltests} top panel).  The result is consistent with null ($p = 0.20$).  The original $T_{\rm LGMCA}$ map also passes this test with $p = 0.08$, indicating that it is already dust-cleaned, though not as thoroughly as $T_{\rm clean}$.

The most stringent dust test is a cross-correlation of $(T_{\rm clean} T_{\rm dust})$ with $\delta_g$ (Fig.~\ref{fig.nulltests} bottom panel).  In our main analysis, we cross-correlate $T_{\rm clean}^2$ with $\delta_g$, while here we replace one factor of $T_{\rm clean}$ with a strong dust tracer.  If $T_{\rm clean}$ contains a significant amount of \emph{WISE}-galaxy-correlated dust, we should see a strong signal here.  The result is roughly consistent with null ($p = 0.02$).  Furthermore, rescaling the approximate amplitude using a standard dust greybody spectrum from $545$ GHz (the $T_{\rm dust}$ template frequency) to the CMB channels that dominate $T_{\rm LGMCA}$ (and $T_{\rm clean}$) at $\approx 100$--$217$ GHz indicates that the dust contribution to the data points in Fig.~\ref{fig.kSZ2xWISE} is $\lesssim 0.003 \, \mu{\rm K}^2$, which is negligible compared to the statistical uncertainties.  A similar test for radio contamination using a CMB-cleaned 30 GHz map yields null ($p = 0.52$).

To further test possible dust leakage due to fluctuating spectral indices amongst the \emph{WISE} galaxies (which might not be fully removed in $T_{\rm clean}$), we construct simulations in which each galaxy is assigned a greybody spectrum with index $\beta$ drawn from a Gaussian of mean $1.75$ and standard deviation $0.06$~\cite{Planck2013CIB}.  The dust temperature is assumed to be $20$ K and the amplitude is set by stacking the 857 GHz map at the \emph{WISE} galaxies' locations (likely an overestimate, since Galactic dust has not been removed).  We generate pairs of maps at various \emph{Planck} frequencies and construct the linear combination that cancels the mean greybody spectrum and preserves the CMB blackbody.  The cross-correlations of these residual maps with $\delta_g$ are a factor $\gtrsim 100$ below the measured $C_{\ell}^{T_{\rm clean}^2 \times \delta_g}$.  These tests establish that any dust leakage in our kSZ${}^2$--\emph{WISE} cross-correlation is significantly below the statistical errors of our measurement.

Finally, we consider different Galactic sky cuts, LGMCA maps from the 2013 \emph{Planck} data, and LGMCA maps with no \emph{WMAP} data, finding results consistent in all cases with our fiducial analysis.

To optimize and test the separation of the CMB lensing and kSZ${}^2$ signals in Fig.~\ref{fig.kSZ2xWISE}, we externally constrain the \emph{WISE} galaxy bias by cross-correlating with \emph{Planck} CMB lensing maps.  Obtaining compatible $b_g$ values from $C_\ell^{T_{\rm clean}^2 \times \delta_g}$ and $C_\ell^{\kappa_{\rm CMB} \delta_g}$ provides a strong test of our approach.  We measure $C_\ell^{\kappa_{\rm CMB} \delta_g}$ in 18 linearly spaced bins over $100 < \ell < 1900$.  We estimate full covariance matrices using 100 \emph{Planck} CMB lensing simulations (separately for the 2013 and 2015 analyses), and include the standard bias correction when computing the inverse covariance matrix~\cite{Hartlapetal2007}.  Fitting the 2015 measurement to a theoretical model based on the most recent ``halofit'' prescription for the non-linear matter power spectrum~\cite{Takahashietal2012}, we find $b_g = 1.13 \pm 0.02$.  We obtain consistent results with the 2013 lensing map (see also~\cite{Ferraroetal2014}) or if we restrict to $100 < \ell < 400$, where theoretical uncertainties due to nonlinearity are significantly diminished.
%%%%%%%%%%%%%%%%%%%%%%%%%%%%%%%%%%%%%%%%%%%%%%

%%%%%%%%%%%%%%%%%%%%%%%%%%%%%%%%%%%%%%%%%%%%%%
\emph{Interpretation---}
We fit a theoretical model consisting of the sum of the $C_\ell^{\rm{kSZ}^2 \times \delta_g}$ prediction in Eq.~(\ref{eq.kSZ2deltagPS}) and the lensing contribution, each with a free amplitude.  The amplitude of the lensing contribution is simply $b_g$.  The amplitude of the kSZ${}^2$ template is $\mathcal{A}_{\rm kSZ^2} b_g$, where $\mathcal{A}_{\rm kSZ^2} = 1$ corresponds to our fiducial model.  We simultaneously fit for $\mathcal{A}_{\rm kSZ^2}$ and $b_g$ assuming a Gaussian likelihood.  We consider three analysis scenarios (Table~\ref{tab:fits}).  Initially, we determine $\mathcal{A}_{\rm kSZ^2}$ and $b_g$ using only $C_\ell^{T_{\rm clean}^2 \times \delta_g}$.  The different shapes of the theoretical templates (see Fig.~\ref{fig.kSZ2xWISE}) allow both amplitudes to be robustly measured.  The kSZ${}^2$ and lensing signals are detected at $3.8\sigma$ and $10\sigma$, respectively.  Marginalizing over $b_g$ only slightly decreases the kSZ${}^2$ $S/N$; if $b_g$ were perfectly known, the kSZ${}^2$ significance would be $4.3\sigma$.  The best-fit model describes the data well, with $\chi^2 = 13.1$ for $11$ degrees of freedom ($p = 0.28$).

The kSZ${}^2$ $S/N$ can be increased by including the external $b_g$ constraint from $C_\ell^{\kappa_{\rm CMB} \delta_g}$.  Including a Gaussian prior centered on $b_g = 1.13$ with standard deviation $0.02$, we find consistent results for $\mathcal{A}_{\rm kSZ^2}$ and $b_g$ compared to the $C_\ell^{T_{\rm clean}^2 \times \delta_g}$-only analysis.\footnote{We neglect the covariance between the $b_g$ information in $C_\ell^{T_{\rm clean}^2 \times \delta_g}$ and $C_\ell^{\kappa_{\rm CMB} \delta_g}$, as the latter dominates the combined constraint.} The kSZ${}^2$ $S/N$ increases to $4.5\sigma$.  Finally, to be conservative, we consider including an additional $10$\% theoretical systematic error on the external $b_g$ prior, due to uncertainties in modeling $\kappa_{\rm CMB}$ and galaxy bias on small scales.  In this case, the $\mathcal{A}_{\rm kSZ^2}$ $S/N$ is $4.2\sigma$.

Using $\mathcal{A}_{\rm kSZ^2} \propto f_{b}^2 f_{\rm free}^2$, the $C_\ell^{T_{\rm clean}^2 \times \delta_g}$-only analysis yields $\left( {f_{b}}/{0.158} \right) \left( {f_{\rm free}}/{1.0} \right) = 1.48 \pm 0.19$ at $z \approx 0.4$.  At this redshift, hydrogen and helium are fully ionized, and thus $f_{\rm free} \approx 1$, with a small fraction of electrons bound in neutral media (e.g., stars or neutral hydrogen gas).  Therefore, our measurement of $f_{b}$ is consistent with the predicted abundance of baryons from the primordial CMB and BBN.\footnote{The posterior $p(f_{b} f_{\rm free})$ is fairly non-Gaussian (see F16); we find $p(f_b f_{\rm free} \leq 1) = 0.054$, equivalent to a 1.6$\sigma$ fluctuation for a Gaussian posterior.} This is the tightest kSZ-derived constraint on $f_{b}$ presented to date.

While our value of $\mathcal{A}_{\rm kSZ^2}$ is slightly high, additional uncertainties must be accounted for in the interpretation (such uncertainties could change the best-fit value of $\mathcal{A}_{\rm kSZ^2}$, but not its detection significance).  The amplitude of $C_\ell^{\rm{kSZ}^2 \times \delta_g} \propto \sigma_8^7$~\cite{Doreetal2004}, where $\sigma_8$ is the matter power spectrum amplitude, and thus a change in $\sigma_8$ within current experimental limits can change the best-fit value of $\mathcal{A}_{\rm kSZ^2}$ at the $\approx 10$\% level.  Also, there are $\approx 5$--$10$\% theoretical uncertainties in the $B_m^{\rm NL}$ fitting function~\cite{Gil-Marinetal2012}, as well as possible non-linear corrections to $v_{\rm rms}^2$, which would affect the inferred $\mathcal{A}_{\rm kSZ^2}$.  Finally, the lensing contamination template is subject to uncertainties at high-$\ell$ due to nonlinear evolution and baryonic physics (if the template shape were highly inaccurate, this could affect the kSZ${}^2$ detection significance, not only the value of $\mathcal{A}_{\rm kSZ^2}$).  However, our comparison to simulations in F16 indicates that the approximations made in our analysis are accurate for both the kSZ${}^2$ and lensing contributions.  We also verify that re-computing the lensing contribution with a $\pm 20$\% shift in the peak of the \emph{WISE} $dn/dz$ (holding the shape fixed) changes the best-fit $\mathcal{A}_{\rm kSZ^2}$ value by $\lesssim 5$\%, well below our statistical errors.

\begin{table}[!h]
\begin{tabular}{l||c|c}
\hline
analysis scenario & $\mathcal{A}_{\rm kSZ^2}$ & $b_g$ \\
\hline
$C_\ell^{T_{\rm clean}^2 \times \delta_g}$ only & $2.18 \pm 0.57$ & $1.10 \pm 0.11$ \\
$C_\ell^{T_{\rm clean}^2 \times \delta_g}$ and $C_\ell^{\kappa_{\rm CMB} \delta_g}$ & $2.24 \pm 0.50$ & $1.13 \pm 0.02$ \\
$+10$\% theory error on $C_\ell^{\kappa_{\rm CMB} \delta_g}$ & $2.21 \pm 0.53$ & $1.11 \pm 0.08$ \\
\hline
\end{tabular}
\caption[]{\label{tab:fits} Fits to the kSZ${}^2$--\emph{WISE} galaxies cross-correlation for three analysis scenarios: (i) using only the $C_\ell^{T_{\rm clean}^2 \times \delta_g}$ data (green circles in Fig.~\ref{fig.kSZ2xWISE}); (ii) including an external constraint on the \emph{WISE} galaxy bias from our measurement of $C_\ell^{\kappa_{\rm CMB} \delta_g}$; (iii) same as (ii), but including an additional $10$\% theoretical systematic error on the $b_g$ constraint from $C_\ell^{\kappa_{\rm CMB} \delta_g}$, due to uncertainties from nonlinear evolution and baryonic physics.}
\end{table}
%%%%%%%%%%%%%%%%%%%%%%%%%%%%%%%%%%%%%%%%%%%%%%

%%%%%%%%%%%%%%%%%%%%%%%%%%%%%%%%%%%%%%%%%%%%%%
\emph{Outlook---}
Our detection confirms that the expected abundance of baryons is present in the low-redshift Universe and that their distribution traces that of the dark matter (an assumption in our model), within the statistical errors.  The novel projected-field estimator implemented here allows kSZ measurements with photometric imaging surveys for the first time, covering much larger sky fractions and tracer samples than spectroscopic data.  In turn, this will yield enormous $S/N$ detections with upcoming multi-frequency CMB surveys (F16), especially for high-resolution experiments that access the kSZ-dominated modes on small angular scales.  We forecast $\gtrsim 100\sigma$ kSZ${}^2$--\emph{WISE} cross-correlations using the upcoming \emph{Advanced ACTPol}~\cite{Hendersonetal2015} and \emph{CMB-S4}~(e.g.,~\cite{Abazajianetal2015}) surveys, contingent on the efficiency of multi-frequency foreground cleaning.  These high-resolution measurements will directly probe the baryon distribution as a function of scale and redshift, and the influence of baryons on the small-scale matter power spectrum.  Moreover, combining kSZ constraints with tSZ analyses will directly determine the gas temperature at the virial radius and beyond.  Finally, kSZ -- large-scale structure cross-correlations will be essential to isolate the high-redshift kSZ signal due to ``patchy'' cosmic reionization from the low-redshift kSZ signal studied here.  This work is a first step toward realizing these exciting new probes of the distribution of gas and matter in our Universe.
%%%%%%%%%%%%%%%%%%%%%%%%%%%%%%%%%%%%%%%%%%%%%%

%%%%%%%%%%%%%%%%%%%%%%%%%%%%%%%%%%%%%%%%%%%%%%
{\small \emph{Acknowledgments.} We are grateful to Olivier Dor\'{e}, Zoltan Haiman, Emmanuel Schaan, Blake Sherwin, Kendrick Smith, and Jessica Werk for useful conversations.  We thank the LGMCA team for publicly releasing their CMB maps.  This work was partially supported by a Junior Fellow award from the Simons Foundation to JCH.  SF was supported in part by the Miller Institute for Basic Research in Science at UC Berkeley.  NB acknowledges support from the Lyman Spitzer Fellowship.  JL is supported by NSF grant AST-1210877.  JCH, SF, and DNS acknowledge support from NSF grant AST1311756 and NASA grant NNX12AG72G.  Some of the results in this paper have been derived using the HEALPix package~\cite{Gorskietal2005}. This publication makes use of data products from the \emph{Wide-field Infrared Survey Explorer}, which is a joint project of the University of California, Los Angeles, and the Jet Propulsion Laboratory/California Institute of Technology, funded by the National Aeronautics and Space Administration.}
%%%%%%%%%%%%%%%%%%%%%%%%%%%%%%%%%%%%%%%%%%%%%%

\end{document}